\documentclass[twocolumn, letter]{aastex631}
\usepackage{amssymb}
\usepackage{xcolor}
\usepackage{colortbl}
\newcommand{\grayline}{\arrayrulecolor{violet}\hline\arrayrulecolor{black}}

\shorttitle{Accurate dust temperature and SFR in the most luminous $z>6$ QSO in the HYPERION sample}
\shortauthors{Tripodi, R. et al.}

\graphicspath{{./}{figures/}}

\begin{document}

\title{Accurate dust temperature and star formation rate in the most luminous $z>6$ quasar in the HYPerluminous quasars at the Epoch of ReionizatION (HYPERION) sample}
\author[0000-0002-9909-3491]{Roberta Tripodi}

\correspondingauthor{R. Tripodi}
\email{roberta.tripodi@inaf.it}

\affiliation{Dipartimento di Fisica, Università di Trieste, Sezione di Astronomia, Via G.B. Tiepolo 11, I-34143 Trieste, Italy}
\affiliation{INAF - Osservatorio Astronomico di Trieste, Via G. Tiepolo 11, I-34143 Trieste, Italy}
\affiliation{IFPU - Institute for Fundamental Physics of the Universe, via Beirut 2, I-34151 Trieste, Italy}

\author[0000-0002-4227-6035]{Chiara Feruglio}
\affiliation{INAF - Osservatorio Astronomico di Trieste, Via G. Tiepolo 11, I-34143 Trieste, Italy}
\affiliation{IFPU - Institute for Fundamental Physics of the Universe, via Beirut 2, I-34151 Trieste, Italy}

\author[0000-0003-2743-8240]{Francisca Kemper}
\affiliation{Institute of Space Science (ICE), CSIC, Can Magrans, E-08193 Cerdanyola del Vallès, Barcelona, Spain}
\affiliation{ICREA, Pg. Lluís Companys 23, E-08010 Barcelona, Spain}
\affiliation{Institut d’Estudis Espacials de Catalunya (IEEC), E-08034 Barcelona, Spain}

\author[0000-0002-2115-1137]{Francesca Civano}
\affiliation{Center for Astrophysics | Harvard \& Smithsonian, Cambridge, MA 02138}

\author[0000-0002-6748-2900]{Tiago Costa}
\affiliation{Max-Planck-Institut für Astrophysik, Karl-Schwarzschild-Straße 1, D-85748 Garching b. München, Germany}

\author[0000-0001-5060-1398]{Martin Elvis}
\affiliation{Center for Astrophysics | Harvard \& Smithsonian, Cambridge, MA 02138}

\author[0000-0002-4314-021X]{Manuela Bischetti}
\affiliation{Dipartimento di Fisica, Università di Trieste, Sezione di Astronomia, Via G.B. Tiepolo 11, I-34143 Trieste, Italy}
\affiliation{INAF - Osservatorio Astronomico di Trieste, Via G. Tiepolo 11, I-34143 Trieste, Italy}

\author[0000-0002-6719-380X]{Stefano Carniani}
\affiliation{Scuola Normale Superiore, Piazza dei Cavalieri 7 I-56126 Pisa, Italy}

\author{Fabio Di Mascia}
\affiliation{Scuola Normale Superiore, Piazza dei Cavalieri 7 I-56126 Pisa, Italy}

\author[0000-0003-3693-3091]{Valentina D'Odorico}
\affiliation{INAF - Osservatorio Astronomico di Trieste, Via G. Tiepolo 11, I-34143 Trieste, Italy}
\affiliation{IFPU - Institute for Fundamental Physics of the Universe, via Beirut 2, I-34151 Trieste, Italy}
\affiliation{Scuola Normale Superiore, Piazza dei Cavalieri 7 I-56126 Pisa, Italy}

\author[0000-0002-4031-4157]{Fabrizio Fiore}
\affiliation{INAF - Osservatorio Astronomico di Trieste, Via G. Tiepolo 11, I-34143 Trieste, Italy}
\affiliation{IFPU - Institute for Fundamental Physics of the Universe, via Beirut 2, I-34151 Trieste, Italy}

\author[0000-0002-7200-8293]{Simona Gallerani}
\affiliation{Scuola Normale Superiore, Piazza dei Cavalieri 7 I-56126 Pisa, Italy}

\author[0000-0002-9122-1700]{Michele Ginolfi}
\affiliation{Dipartimento di Fisica e Astronomia, Università di Firenze, Via G. Sansone 1, 50019, Sesto Fiorentino (Florence), Italy}
\affiliation{INAF - Osservatorio di Arcetri, Largo E. Fermi 5, I-50125, Florence, Italy}

\author[0000-0002-4985-3819]{Roberto Maiolino}
\affiliation{Institute of Astronomy, University of Cambridge, Madingley Road, Cambridge CB3 0HA, UK}
\affiliation{Kavli Institute for Cosmology, University of Cambridge, Madingley Road, Cambridge CB3 0HA, UK}
\affiliation{Department of Physics and Astronomy, University College London, Gower Street, London WC1E 6BT, UK}

\author[0000-0001-9095-2782]{Enrico Piconcelli}
\affiliation{INAF - Osservatorio Astronomico di Roma, Via Frascati 33, I-00040 Monte Porzio Catone, Italy}

\author[0000-0003-3050-1765]{Rosa Valiante}
\affiliation{INAF - Osservatorio Astronomico di Roma, Via Frascati 33, I-00040 Monte Porzio Catone, Italy}

\author[0000-0002-4205-6884]{Luca Zappacosta}
\affiliation{INAF - Osservatorio Astronomico di Roma, Via Frascati 33, I-00040 Monte Porzio Catone, Italy}

\date{Accepted in ApJL on March 20, 2023}

\begin{abstract}
We present ALMA Band 9 continuum observation of the ultraluminous quasi-stellar object (QSO) SDSS J0100+2802, providing a $\sim 10\sigma$ detection at $\sim 670$ GHz. SDSS J0100+2802 is the brightest QSO with the most massive super massive black hole (SMBH) known at $z>6$, and we study its dust spectral energy distribution in order to determine the dust properties and the star formation rate (SFR) of its host-galaxy. We obtain the most accurate estimate so far of the temperature, mass and emissivity index of the dust, having $T_{\rm dust}=48.4\pm2.3$ K, $M_{\rm dust}=(2.29\pm0.83)\times 10^7$ M$_\odot$, $\beta=2.63\pm 0.23$. This allows us to measure the SFR with the smallest statistical error for this QSO, SFR$=265\pm 32\ \rm M_\odot yr^{-1}$. Our results enable us to evaluate the relative growth of the SMBH and host galaxy of J0100+2802, finding that the SMBH is dominating the process of BH-galaxy growth in this QSO at $z=6.327$, when the Universe was $865$ Myr old. 
 Such unprecedented constraints on the host galaxy SFR and dust temperature can only be obtained through high frequency observations, and highlight the importance of ALMA Band 9 to obtain a robust overview of the build-up of the first quasars' host galaxies at $z>6$.
\end{abstract}

\vspace{-2cm}

\section{Introduction}

In the past decade, the Atacama Large Millimeter/sub-millimeter Array (ALMA), along with the Northern Extended Millimeter Array (NOEMA), the Very Large Array (VLA), and \textit{Herschel}, have probed the cold gas and dust of quasi-stellar-object (QSO) host galaxies. The dust continuum was detected in many $z \sim 6$ QSOs, with far-infrared (FIR) luminosities of $L_{\rm FIR}=10^{11-13}\ \rm L_{\odot}$ and dust masses of about $M_{\rm dust}= 10^{7-9}\ \rm M_{\odot}$ \citep{decarli2018, carniani+19, shao2019}. The rest-frame FIR continuum emission originates from dust heated by the ultraviolet (UV) radiation from young and massive stars \citep{decarli2018, venemans2020, neeleman2021} and the active galactic nuclei (AGN) radiation \citep{schneider2015, dimascia2021, walter2022}. The latter contribution is usually neglected when modelling the FIR spectral energy distribution (SED) of $z\sim 6$ QSOs, although the AGN heating can contribute $30-70$\% of the FIR luminosity \citep{schneider2015, duras2017}. Moreover, dust masses are often determined with huge uncertainties relying only on single-frequency continuum detections. However, if multi-frequency ALMA observations are available in the rest-frame FIR, probing both the peak and the Rayleigh Jeans tail of the dust SED, the dust temperature and mass can be constrained with statistical uncertainties $<10-20\%$ (e.g. \citealt{carniani+19, Tripodi2022}), resulting into an high accuracy in the determination of the star formation rate (SFR).

In this Letter we present ALMA Band 9 observations of the QSO SDSS J010013.02+280225.8 (hereafter J0100+28) at $z_{\rm [CII]}=6.327$ \citep{wang2019}. \citet{wu2015} estimated a bolometric luminosity of $L_{\rm bol}=4.29\times 10^{14}\rm \ L_{\odot}$ and a BH mass of $M_{\rm BH}=1.24\times10^{10}\ \rm M_{\odot}$ for J0100+28, making it the most optically luminous QSO with the most massive SMBH known at $z>6$. Both measurements have been recently confirmed by JWST \citep{eilers2022}. \citet{wang2019} performed a multi-frequency analysis of the dust SED, but they could not obtain a precise determination of the dust properties, concluding that J0100+28 has either a high dust emissivity ($\beta \gtrsim 2$) or a high dust temperature ($T_{\rm dust} \gtrsim 60$ K), or a combination of thereof.

J0100+28 belongs to the HYPerluminous quasars at the Epoch of ReionizatION (HYPERION) sample of $z>6$ QSOs, selected to include the QSOs with the most massive SMBH at their epoch, possibly resulting from an exceptional fast mass growth during their accretion history. The sample consists of 17 QSOs with $L_{\rm bol} = 10^{47.3}$ erg/s, SMBH masses in the range $M_{\rm BH} = 10^{9}-10^{10}\ \rm M_\odot$, and Eddington ratios $>0.3-0.4$. The HYPERION sample and the details of its selection will be presented in Zappacosta et al. (in prep.).

The goal of this work is to derive the most accurate estimate of the dust mass, dust temperature, and therefore of the SFR for the host galaxy of J0100+28. 
We adopt a $\Lambda$CDM cosmology from \citet{planck2018}: $H_0=67.4\ \rm km\ s^{-1}\ Mpc^{-1}$, $\Omega_m = 0.315$ and $\Omega_{\Lambda} = 0.685$. Thus, the angular scale is $5.66$ kpc/arcsec at $z=6.3$.

\section{Observation}
We analyse the dataset 2021.2.00151.S from the ALMA 7m array, designed to detect the continuum emission at a frequency of 670.91 GHz in Band 9, and has a total integration time of 2.2 hours. The visibility calibration and imaging are performed through the Common Astronomy Software Applications (CASA; \citealt{mcmullin2007}), version 5.1.1-5. 
We apply \texttt{tclean} using natural weighting and a $3\sigma$ cleaning threshold.
We image the continuum by collapsing all channels, selected by inspecting the visibilities in all spectral windows. 
We obtain a clean beam of ($1.97 \times 1.17$) arcsec$^2$, corresponding to a spatial resolution of $\sim11$ kpc, and an r.m.s. noise of 0.8 mJy/beam in the continuum.

\section{Analysis}
\label{sec:analysis}

\subsection{QSO continuum emission and dust properties}
\label{sec:sed}
\begin{figure}
	\centering
	\includegraphics[width=0.9\linewidth]{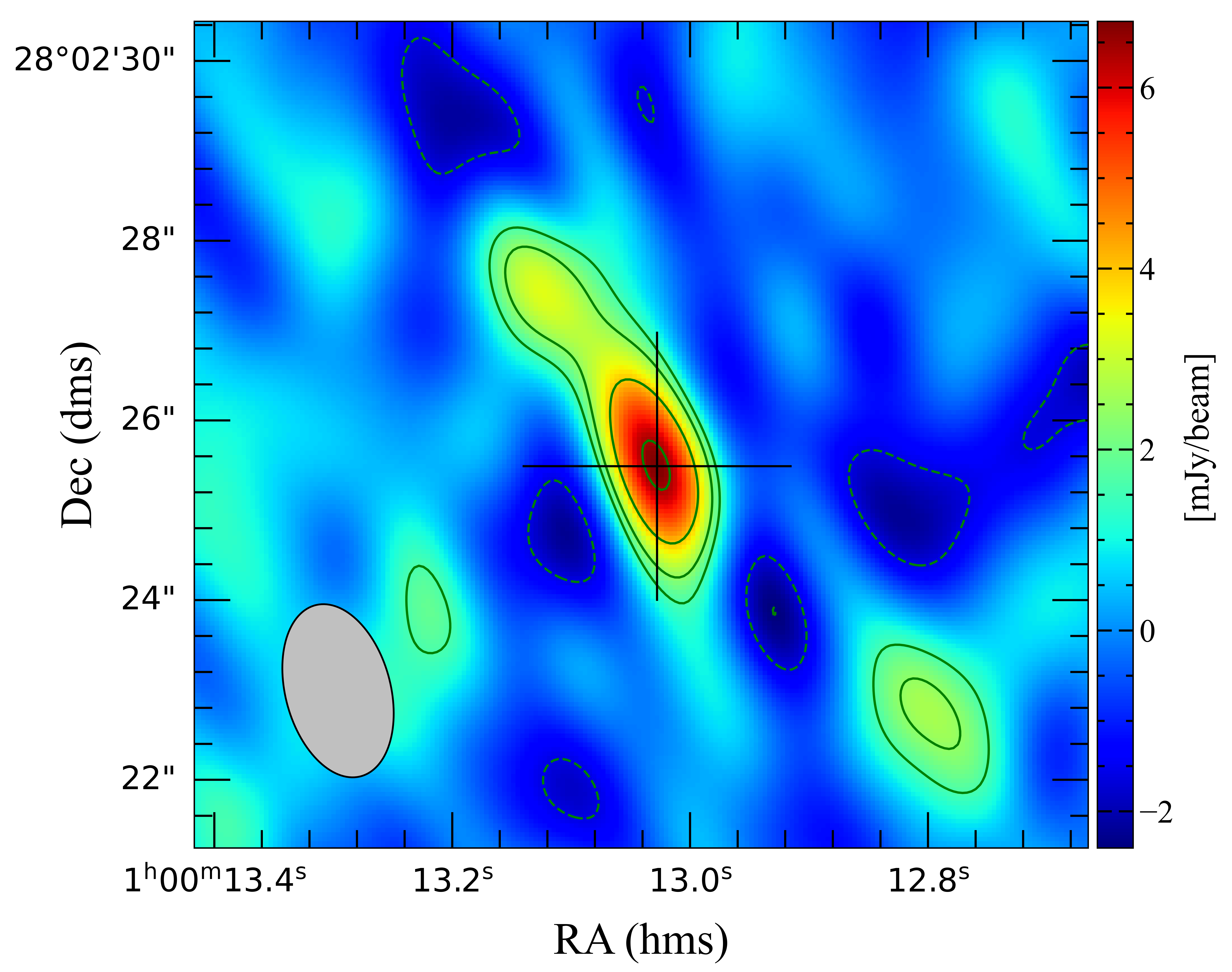}
\caption{\footnotesize 670.9 GHz dust continuum map of QSO J0100+28 (levels $-3,-2,2,3,5,\text{and }8\sigma$, $\sigma = 0.8$ mJy/beam). The clean beam ($1.97\times 1.17\rm \ arcsec^2$) is indicated in the lower left corner of the diagram. The cross indicates the position of the continuum peak.}
\label{fig:cont}
\end{figure}

Figure \ref{fig:cont} presents the 670.9 GHz continuum emission map of J0100+28. The peak flux density is $6.99 \pm 0.71$ mJy/beam. The source is not spatially resolved. We check that the other FIR and radio flux measurements were extracted from a region similar to our resolution\footnote{\citet{wang2019} used tapered maps at 1.5 arcsec in order not to miss the fainter extended emission. \citet{liu2022} also tapered their 6 and 10 GHz maps, that had a resolution of $\sim 1.5$ arcsec, to match the resolution of the 1.5 GHz map ($\sim 4$ arcsec), and they do not find significant differences in flux density between the tapered and the full-resolution maps.}. 

\begin{figure*}
	\centering
	\includegraphics[width=1\linewidth]{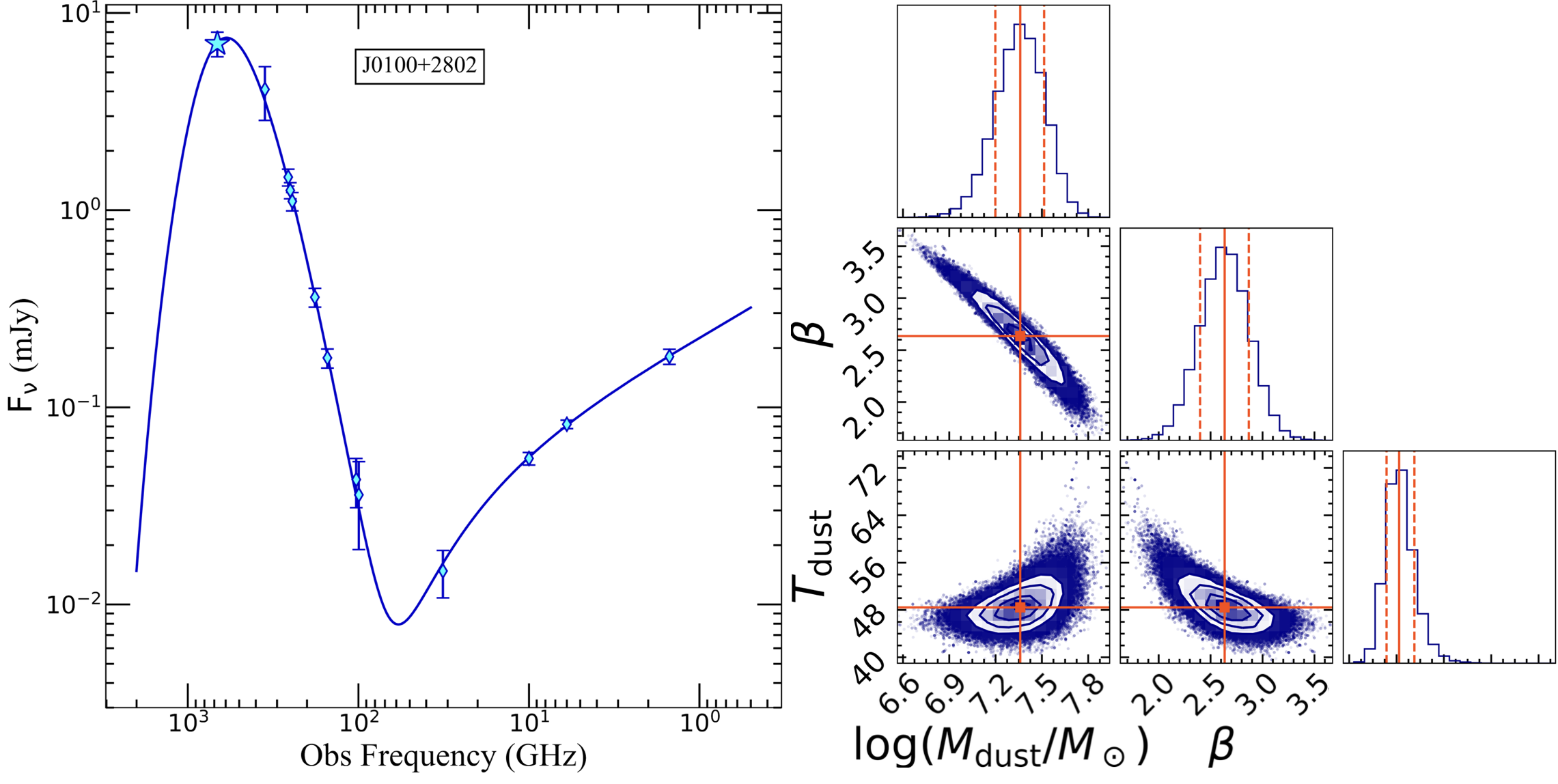}
\caption{\footnotesize Left panel: SED of J0100+28 using our new ALMA 670.91 GHz data (cyan star), the continuum fluxes from 32 GHz to 353 GHz \citep{wang2019}, and at 1.5, 6 and 10 GHz \citep{li2022} (cyan diamonds). The best-fitting curve is shown as a blue solid line. Right panel: Corner plot showing the posterior probability distributions of $T_{\rm dust}, M_{\rm dust}, \beta$. Orange solid lines indicate the best-fitting value for each parameter, while the dashed lines mark the 16th and 84th percentiles for each parameter.}
\label{fig:sed}
\end{figure*}

We perform an SED fitting using the continuum emission measured at $\sim 671$ GHz, together with the emissions presented in \citet{wang2019} from 32 GHz to 353 GHz, and in \citet{liu2022} at 1.5, 6 and 10 GHz. Although we are interested in the cold dust properties of the QSO host galaxy, such as $T_{\rm dust}$, $M_{\rm dust}$ and $\beta$, we consider also the contribution of the lower frequency emission to the dust SED, since in general it may not be negligible. \citet{liu2022} noticed a time variability among their and previous measurements of the radio continuum in the range [6-10] GHz. For the sake of simplicity, we consider the most recent measurements of the radio continuum emission (i.e., those from \citealt{liu2022}). However, we verified that our results do not change by considering all the measurements available in the radio band. 

We model the dust continuum with a modified black-body (MBB) function and the low frequency radio emission using a power law with an exponential cut-off (PLCO). Details about the fitting functions and procedure can be found in the Appendix \ref{app:a}. The model has six fitting parameters: dust temperature ($T_{\rm dust}$), dust mass ($M_{\rm dust}$), dust emissivity index ($\beta$) entering in the MBB function, and the normalization ($n$), the radio power-law spectral index ($\alpha$), and the cut-off frequency ($\nu_{\rm cutoff}$) for the PLCO. It is worth to stress that the galaxy is likely characterized by a distribution of dust temperatures, that may reach hundreds of K close to the AGN (see e.g., \citealt{walter2022}) and may decreases toward the outskirts, hence the temperature derived from the SED fitting should be interpreted as an `effective' dust temperature\footnote{We adopt the term `effective temperature' since we are not able to map the spatial distribution of the dust temperatures across the galaxy, similarly to the definition of the effective temperature of a star.}. The best-fit model has $T_{\rm dust}=48.4\pm2.3$ K, $M_{\rm dust}=(2.29\pm0.83)\times 10^7$ M$_\odot$, $\beta=2.63\pm 0.23$, $n=0.08\pm 0.01$ mJy, $\alpha=0.48\pm 0.09$ and $\nu_{\rm cutoff}=235 \pm 100$ GHz. Figure \ref{fig:sed} shows the observed SED fitted by our best-fit model (left panel) and the posterior distributions for the dust parameters (right panel). Posterior distributions of all parameters is reported in Appendix \ref{app:a}. We find a gas-to-dust ratio ${\rm GDR}=236\pm155$ based on our $M_{\rm dust}$ estimate and the molecular gas mass obtained by \citet{wang2019} (see Table \ref{tab:properties}). This is in agreement with GDRs found in ultraluminous QSOs at $z\sim 2-4$ \citep{bischetti2021}, and in local galaxies at solar metallicities \citep{devis2019}. This latter comparison would imply that J0100+28's host galaxy has already been highly enriched with metals. 

We estimate the total infrared (TIR) luminosity for the best-fit model by integrating from $8$ to $1000\ \mu$m rest-frame, obtaining $L_{\rm TIR} = 5.30\pm 0.64\times 10^{12}\ \rm L_{\odot}$. This would imply a SFR of $530\pm 64\ \rm M_{\odot}\ yr^{-1}$, adopting a Chabrier initial mass function \citep{chabrier2003}. However, several observations and radiative transfer simulations suggested that the radiative output of luminous QSOs substantially contributes to dust heating on kpc scale \citep{schneider2015, dimascia2021, walter2022}. \citet{duras2017} showed that in average $\sim 50\%$ of the total IR luminosity in QSOs with $L_{\rm bol}>10^{47}\rm \ erg ~s^{-1}$ is due to dust heated by the AGN radiation. Recently, \citet{dimascia2022} found a correction factor of $1/30$ for the SFR of the brightest object in their sample. However, the $L_{\rm bol}$ of J0100 is above the range explored in their simulations and, moreover, J0100 has very peculiar properties in terms of UV magnitude and dust temperatures with respect to the simulated QSOs in \citet{dimascia2022}. Therefore, we consider the average correction proposed by \citet{duras2017} as the most appropriate choice, and we obtain ${\rm SFR} = 265\pm 32\ \rm M_{\odot}yr^{-1}$.

\begin{table}
\vspace{0.2cm}
		\caption{Properties of SDSS J0100+2802}
		\centering
		\begin{tabular}{llc}
			\hline
			\hline
			$T_{\rm dust}$ & [K] & $48.4 \pm 2.3$\\[0.1cm]
			$M_{\rm dust}$ & $[10^{7}\ \rm M_{\odot}]$ &  $2.29\pm 0.83$\\[0.1cm]
			$\beta$ & & $2.63 \pm 0.23$\\[0.1cm]
			SFR$^{\rm (a)}$ & [$\rm M_{\odot}\ yr^{-1}$] & $265\pm 32$\\[0.1cm]
			GDR & & $236\pm 155$ \\[0.1cm]
			\grayline
			$M_{\rm gas}^{\rm (b)}$ & $[10^{10}\ \rm M_{\odot}]$ & $0.54\pm 0.16$\\[0.1cm]
			$M_{\rm dyn}$(CV)$^{\rm (c)}$ & $ [10^{10}\ \rm M_{\odot}]$ &  $\sim 890$\\[0.1cm]
			$M_{\rm dyn}$(VT)$^{\rm (d)}$ & $ [10^{10}\ \rm M_{\odot}]$ &  $3.25\pm 0.46$\\[0.1cm]
			$M_{\rm BH}^{\rm (e)}$ & $[10^{10}\ \rm M_{\odot}]$ &  $1.05$  \\ [0.1cm]
			\hline
		\end{tabular}
		\label{tab:properties}
		\flushleft 
		\footnotesize {{\bf Notes.} Quantities above the violet line are from this work, while the other are taken from \citet{wang2019}. $^{\rm (a)}$ The SFR is corrected by a factor of 50\%, accounting for the contribution of the QSO. $^{\rm (b)}$ The gas mass is estimated within a diameter of $\sim 1.4$ kpc \citep{wang2019}.$^{\rm (c)}$ Dynamical mass estimated using the circular velocity assuming a disk inclination of $5^{\circ}$ within 1.8 kpc radius \citep{wang2019}. $^{\rm (d)}$ Dynamical mass estimated using the virial theorem within 1.8 kpc radius \citep{wang2019}. $^{\rm (e)}$ BH mass from MgII emission line (Mazzucchelli et al in prep.).} 
	\end{table}
\section{Discussion and Conclusions}

\begin{figure*}
    \vspace{0.5cm}
	\centering
	\includegraphics[width=1\linewidth]{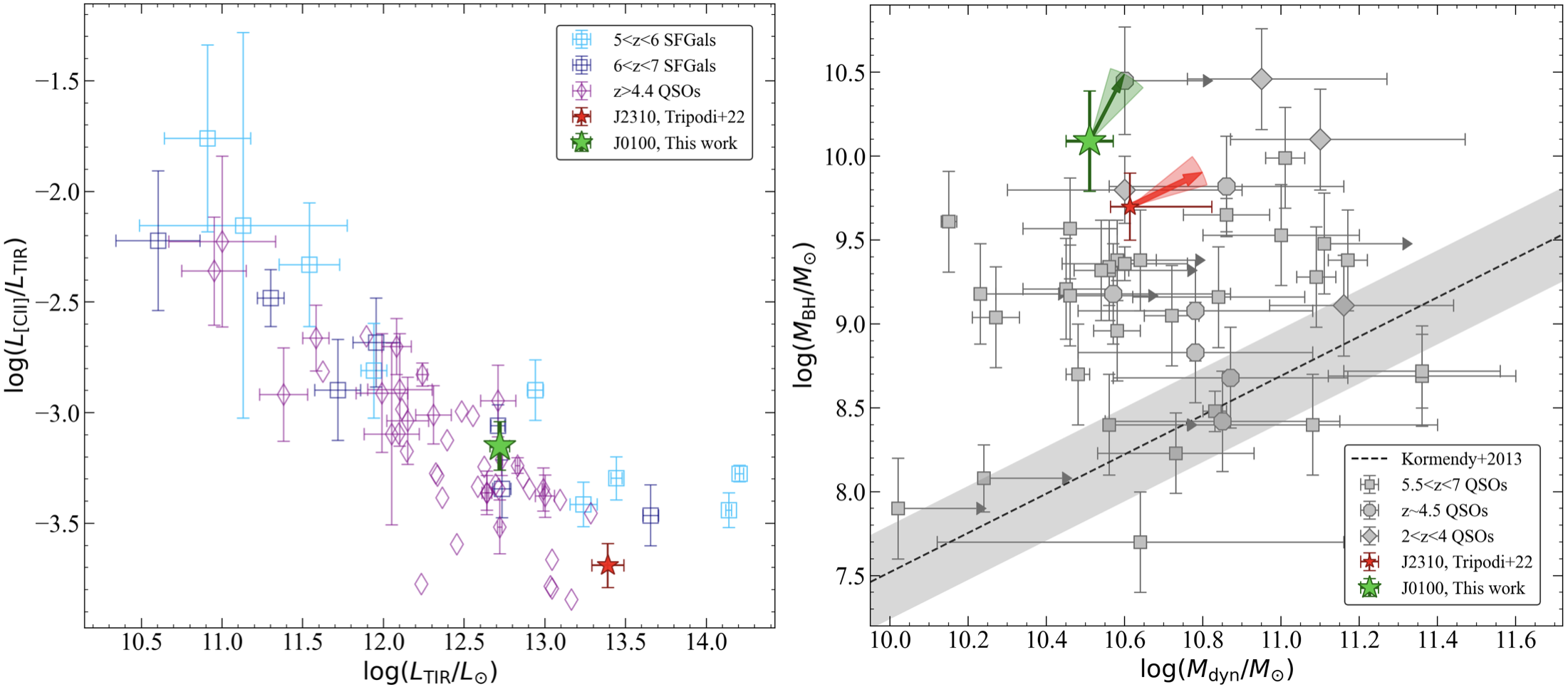}
\caption{\footnotesize Left panel: [CII]-to-TIR luminosity ratio vs TIR luminosity. We show our result for QSO J0100+28 (green star) compared with QSO J2310+18 at $z=6.0028$ (red star, see text), two samples of star-forming galaxies at $5<z<6$ and $6<z<7$ (cyan and blue squares respectively, taken from \citealt{lagache2018}), a sample of QSOs at $z>4.4$ (violet diamonds, from \citealt{bischetti2021} and \citealt{decarli2018}). Right panel: \footnotesize BH mass vs dynamical mass for J0100+28 (green star), compared with QSO J2310+18 at $z=6.0028$ (red star, \citealt{Tripodi2022}), WISSH QSOs at $z\sim 2-4$ (grey diamonds, \citealt{bischetti2021}), and luminous $z\sim 4-7$ QSOs (grey dots and grey squares, \citealt{venemans2016,venemans2017a,willott2013,willott2015,willott2017,kimball2015,trakhtenbrot2017, feruglio2018, mortlock2011, derosa2014,kashikawa2015, neeleman2021}). Black dashed line (and shaded area) is the local relation from \citet{kormendy2013}. For J0100+28 (J2310+18), the slope of the arrow, with its uncertainty, indicates how much the growth efficiency of the SMBH is increasing (slowing down) with respect to the growth of the host galaxy.}
\label{fig:lcii-ltir}
\end{figure*}

By fitting the dust SED of the QSO J0100+28, we found that our observation in Band 9 favours a dust temperature lower than  the effective dust temperatures found in simulations in bright ($L_{\rm bol} > 10^{13}\ \rm L_\odot$) quasar-hosts ($\sim 90$ K, e.g. \citealt{dimascia2022}). This discrepancy can be due to different dust spatial distributions between the simulated objects and J0100+28, or to limits in the dust modelling and radiative transfer post-processing (e.g. the absence of a dusty torus in \citealt{dimascia2021}). With the current unresolved observation we are not able neither to constrain different temperature components, nor to determine the temperature distribution across the galaxy.
The value found for the emissivity index is higher than $\beta=1.6$ found for the average SED of high-z QSOs  \citep{beelen+2006}. However, as noted by \citet{beelen+2006}, large variations of $\beta$ are found when considering individual QSO SED. 
Only excluding the Band 9 data, we would obtain a good fit with $\beta<2$, having then a factor of two higher temperature \citep[see also][]{wang2019}. This shows that the results can be misguided by relying only on the lowest frequency data points, while Band 9 is essential to reliably estimate dust parameters up to the highest redshifts. Indeed, \citet{novak2019}, using observations up to Band 8 at $\sim$404 GHz, could not constrain $T_{\rm dust}$ in the $z=7.54$ quasar ULAS J1342+0928. This work provides the first measurement of the Band 9 continuum of a luminous QSO host galaxy. Similarly, \citet{bakx2021} used Band 9 observations to constrain $T_{\rm dust}$ for a galaxy at $z=7.13$.

It is not straightforward to assess the physical reasons of the high $\beta$ value. In principle, $\beta$ depends on the physical properties and chemical composition of the grains, and possibly on environment and temperature. There are cases in which $\beta$ can be larger than 2 \citep[see e.g.,][]{valiante2011, galliano2018}. Spatially resolved studies in low-z galaxies showed a spread of $\beta$ within a single galaxy, probably due to the temperature mixing or the different properties of the grain populations, or both. The strong anti-correlation between $T_{\rm dust}$ and $\beta$ can arise from the MCMC fitting procedure itself, mainly if there is a massive amounts of cold dust (i.e., $T\lesssim 10$ K, \citealt{galliano2018}). However, in our case, the accurate sampling of the SED from low to high frequency significantly relaxed the strength of $T_{\rm dust}-\beta$ anti-correlation (Figure \ref{fig:sed}). Therefore, we are confident that the estimates of $\beta$ and $T_{\rm dust}$ are not strongly biased due to the effect of this anti-correlation. Finding a physical explanation for the high value of $\beta$ would require studies of the properties of the dust grains and/or of a detailed analysis of the temperature mixing and is beyond the scope of this paper. 

 The left panel of Figure \ref{fig:lcii-ltir} shows the ratio of integrated [CII] to TIR luminosity for J0100+28 and a compilation of high redshift QSOs and galaxies (see caption).
For J0100+28 we find $L_{\rm [CII]}/L_{\rm TIR}\sim 7\times 10^{-4}$, given $L_{\rm [CII]}=3.7\times 10^{9}\ \rm L_\odot$ \citep{wang2019}. This [CII] deficit is also predicted for high-z galaxies by semi-analytical models of galaxy evolution (e.g.  \citet{lagache2018}), where  the [CII] deficit arises from the high intensity of the interstellar radiation field. Our estimate of $L_{\rm [CII]}/L_{\rm TIR}$  agrees well with their results at $z\sim 6$ when we extrapolate their predictions at higher $L_{\rm TIR}$. \citet{carniani2018} found that the local $L_{\rm [CII]}-{\rm SFR}$ relation for star-forming galaxies (see \citealt{delooze2014}) is still valid at high-z, but with a dispersion twice higher than observed locally. Our results agree within $<1\sigma$ with the correlation of \citet{carniani2018}, and with the results for high-z galaxies of \citet{lagache2018}.


We evaluate the evolutionary state of the SMBH - host galaxy system. 
\citet{wang2019} provides two estimates for $M_{\rm dyn}$, one based on the assumption that the [CII] line arises from a rotating disk with very low inclination, and the other using the virial theorem with the hypothesis that the gas is supported by random motion (see Table \ref{tab:properties}). Although both estimates carry large uncertainties, the latter is in agreement with the dynamical masses commonly measured for other QSOs at high-z (see right panel of Figure \ref{fig:lcii-ltir}) and we adopt it as a lower limit.
The molecular gas mass is $M_{\rm gas}=5.4\times 10^9\ \rm M_\odot$, and this implies a molecular gas fraction $\mu=M_{\rm gas}/M_{*}=M_{\rm gas}/(M_{\rm dyn}-M_{\rm BH}-M_{\rm gas})=0.4$, which is remarkably lower than the typical gas fractions at $z\sim 3-4$ \citep{tacconi2018}. 
We define the growth efficiency of the galaxy as ${\rm SFR}/M_{\rm g+s}$, where $M_{\rm g+s}=M_{\rm gas}+M_{*}$  and the SFR is corrected for the QSO contribution. We use $M_{\rm g+s}$, instead of $M_{\rm dyn}$ (as in \citealt{Tripodi2022}), since the BH mass is not negligible ($M_{\rm BH}\sim 40\% M_{\rm dyn}$, \citealt{wang2019}) and can bias our results\footnote{To be consistent, we re-compute the growth efficiency of the galaxy for J2310 using $M_{\rm g+s}$, and we changed the slope of the arrow with the respect to the one shown in \citet{Tripodi2022}. However, the results do not change significantly, since $M_{\rm g+s}\simeq M_{\rm dyn}$ for J2310.}.  We obtain $M_{\rm g+s}=2.01\times 10^{10}\ \rm M_{\odot}$, therefore ${\rm SFR}/M_{\rm g+s}=1.3 \times 10^{-8}\rm ~yr^{-1}$. 
On the other hand, we derive a BH growth efficiency\footnote{$\dot M_{\rm BH}=L_{\rm bol}/(\epsilon ~c^2)$, where $\epsilon$ is the radiative efficiency, and $c$ is the speed of light.}, $(1-\epsilon) \dot M_{\rm BH}/M_{\rm BH}=2.5\times 10^{-8} \rm ~yr^{-1}$, where we use the BH mass derived from  MgII (Table \ref{tab:properties}), and assume a radiative efficiency $\epsilon=0.1$ (e.g. \citealt{marconi2004}). The right panel of Figure \ref{fig:lcii-ltir} shows $M_{\rm BH}$ versus $M_{\rm dyn}$ for J0100+28, J2310+18 \citep{Tripodi2022} and a compilation of QSOs at different redshifts, comparing them with the local $M_{\rm BH}-M_{\rm dyn}$ relation. The majority of QSOs, including J0100+28, lie above the local relation in the BH dominance regime \citep{volonteri2012}. However, this tension can be partially softened if accounting for the uncertainties on the dynamical mass estimates. These mainly depend on the determination of the disk inclination and can be significantly high for some QSOs \citep{valiante2014, pensabene2020}. This is also the case of J0100+28, whose $M_{\rm dyn}$ can be in principle as high as $10^{12}\ \rm M_\odot$ (see Table \ref{tab:properties}). For J0100+28, we found $(1-\epsilon)\dot M_{\rm BH}/M_{\rm BH}>{\rm SFR}/M_{\rm g+s}$, suggesting that the BH is still dominating the process of BH-galaxy growing in this QSO at $z=6.327$. This result is still valid if considering lower SFR (adopting the correction by \citealt{dimascia2022}) and/or higher $M_{\rm dyn}$ (i.e. higher $M_{\rm g+s}$ at fixed BH mass), since the galaxy growth factor would be even smaller. Our results do not consider the gas inflow. However, we expect this term to contribute on average to both SFR and $M_{\rm g+s}$, leaving their ratio mostly unaffected. On the other hand, in QSO J2310+18 at $z\sim6$ AGN feedback might be slowing down the accretion onto the SMBH, while the host galaxy grows fast \citep{Tripodi2022,bischetti2022}. The different evolutionary state of J0100+28 and J2310+18, separated by only $\sim 60$ Myr (i.e. $\Delta z\sim 0.3$),  arise mainly from the difference in their BH mass (i.e., $M_{\rm BH, J0100+28}\sim 2\times\ M_{\rm BH, J2310+18}$) and in SFR (i.e., ${\rm SFR_{J0100+28}}\sim 0.2\times {\rm SFR_{J2310+18}}$). 
In principle,  ${\rm SFR}/M_{*}$ is a better probe of the galaxy growth, however this is not available for most high redshift QSOs. For J0100+28, using ${\rm SFR}/M_*$ would not affect our results since $M_{\rm g+s} \approx M_*$. For J2310, $M_{*}\ll M_{\rm g+s}$, therefore resulting in an even flatter slope. 
 The BH dominated growth of J0100+28 matches the qualitatively expectations for the evolutionary state of the HYPERION QSOs, that were selected to be the most luminous QSOs with the most massive SMBH at their epochs.

We compare our results with `zoom-in' simulations of QSOs using the moving-mesh code AREPO, following BH growth and feedback via energy-driven outflows \citep{costa2014a,costa2015}. We found that simulations reproduce BHs only up to masses $\sim 10^9\ \rm M_\odot$ that have host galaxies with dynamical masses $\sim 10^{11}\ \rm M_\odot$. These are considerably more massive than J0100+28 host-galaxy. The growth of the system is characterised by intermittent phases, where the BH and galaxy-dominated growth phases change on short timescales. Hence, the diagnostic power of the relation $(1-\epsilon)\dot M_{\rm BH}/M_{\rm BH}$ - ${\rm SFR}/M_{\rm g+s}$ needs to be validated in larger samples that we plan to investigate in a forthcoming work.

In summary, this work allowed us to measure the SFR with the smallest statistical error, reaching an accuracy of $\sim 15\%$ for the QSO J0100+28. Such unprecedented constraints on the host galaxy SFR (and $T_{\rm dust}$) highlight the critical role of ALMA Band 9 to obtain a robust overview of the build-up of SMBHs and their massive host galaxies at the epoch of Reionisation. The systematic uncertainty on the SFR is still high due to the uncertainty in the estimate of the QSO contribution. Recently, \citet{tsukui2023} estimated this contribution for a $z=4.4$ QSO using high resolution ALMA observations up to band 9. Given our current unresolved observation, we are not able to use the same approach. However, we plan to determine this correction factor in forthcoming works using both high resolution observations and radiative transfer models to reproduce the observed SED.

\begin{acknowledgments}
\textit{Acknowledgments}. We thank the anonymous referee for the comments and suggestions that helped us improving this work. This paper makes use of the following ALMA data: ADS/JAO.ALMA\#2021.2.00151.S. ALMA is a partnership of ESO (representing its member states), NFS (USA) and NINS (Japan), together with NRC (Canada), MOST and ASIAA (Taiwan) and KASI (Republic of Korea), in cooperation with the Republic of Chile. The Joint ALMA Observatory is operated by ESO, AUI/NRAO and NAOJ. RT acknowledges financial support from the University of Trieste. RT, CF, FF, MB acknowledge support from PRIN MIUR project “Black Hole winds and the Baryon Life Cycle of Galaxies: the stone-guest at the galaxy evolution supper”, contract \#2017PH3WAT. FK acknowledges the Spanish program Unidad de Excelencia María de Maeztu CEX2020-001058-M, financed by MCIN/AEI/10.13039/501100011033. SC is supported by the European Union (ERC, WINGS,101040227). RM acknowledges ERC Advanced Grant 695671 QUENCH, and support from the UK Science and Technology Facilities Council (STFC). RM also acknowledges funding from a research professorship from the Royal Society.
\end{acknowledgments}

\vspace{5mm}
\facilities{ALMA}

\software{astropy, CASA (v5.1.1-5, \citealt{mcmullin2007})}
          

\begin{thebibliography}{}
	\expandafter\ifx\csname natexlab\endcsname\relax\def\natexlab#1{#1}\fi
	\providecommand{\url}[1]{\href{#1}{#1}}
	\providecommand{\dodoi}[1]{doi:~\href{http://doi.org/#1}{\nolinkurl{#1}}}
	\providecommand{\doeprint}[1]{\href{http://ascl.net/#1}{\nolinkurl{http://ascl.net/#1}}}
	\providecommand{\doarXiv}[1]{\href{https://arxiv.org/abs/#1}{\nolinkurl{https://arxiv.org/abs/#1}}}
	
	\bibitem[{{Bakx} {et~al.}(2021){Bakx}, {Sommovigo}, {Carniani}, {Ferrara},
		{Akins}, {Fujimoto}, {Hagimoto}, {Knudsen}, {Pallottini}, {Tamura}, \&
		{Watson}}]{bakx2021}
	{Bakx}, T. J.~L.~C., {Sommovigo}, L., {Carniani}, S., {et~al.} 2021, \mnras,
	508, L58, \dodoi{10.1093/mnrasl/slab104}
	
	\bibitem[{{Beelen} {et~al.}(2006){Beelen}, {Cox}, {Benford}, {Dowell},
		{Kov{\'a}cs}, {Bertoldi}, {Omont}, \& {Carilli}}]{beelen+2006}
	{Beelen}, A., {Cox}, P., {Benford}, D.~J., {et~al.} 2006, \apj, 642, 694,
	\dodoi{10.1086/500636}
	
	\bibitem[{{Bischetti} {et~al.}(2021){Bischetti}, {Feruglio}, {Piconcelli},
		{Duras}, {P{\'e}rez-Torres}, {Herrero}, {Venturi}, {Carniani}, {Bruni},
		{Gavignaud}, {Testa}, {Bongiorno}, {Brusa}, {Circosta}, {Cresci},
		{D'Odorico}, {Maiolino}, {Marconi}, {Mingozzi}, {Pappalardo}, {Perna},
		{Traianou}, {Travascio}, {Vietri}, {Zappacosta}, \& {Fiore}}]{bischetti2021}
	{Bischetti}, M., {Feruglio}, C., {Piconcelli}, E., {et~al.} 2021, \aap, 645,
	A33, \dodoi{10.1051/0004-6361/202039057}
	
	\bibitem[{{Bischetti} {et~al.}(2022){Bischetti}, {Feruglio}, {D'Odorico},
		{Arav}, {Ba{\~n}ados}, {Becker}, {Bosman}, {Carniani}, {Cristiani}, {Cupani},
		{Davies}, {Eilers}, {Farina}, {Ferrara}, {Maiolino}, {Mazzucchelli},
		{Mesinger}, {Meyer}, {Onoue}, {Piconcelli}, {Ryan-Weber}, {Schindler},
		{Wang}, {Yang}, {Zhu}, \& {Fiore}}]{bischetti2022}
	{Bischetti}, M., {Feruglio}, C., {D'Odorico}, V., {et~al.} 2022, \nat, 605,
	244, \dodoi{10.1038/s41586-022-04608-1}
	
	\bibitem[{{Carniani} {et~al.}(2018){Carniani}, {Maiolino}, {Amorin},
		{Pentericci}, {Pallottini}, {Ferrara}, {Willott}, {Smit}, {Matthee},
		{Sobral}, {Santini}, {Castellano}, {De Barros}, {Fontana}, {Grazian}, \&
		{Guaita}}]{carniani2018}
	{Carniani}, S., {Maiolino}, R., {Amorin}, R., {et~al.} 2018, \mnras, 478, 1170,
	\dodoi{10.1093/mnras/sty1088}
	
	\bibitem[{{Carniani} {et~al.}(2019){Carniani}, {Gallerani}, {Vallini},
		{Pallottini}, {Tazzari}, {Ferrara}, {Maiolino}, {Cicone}, {Feruglio}, {Neri},
		{D'Odorico}, {Wang}, \& {Li}}]{carniani+19}
	{Carniani}, S., {Gallerani}, S., {Vallini}, L., {et~al.} 2019, \mnras, 489,
	3939, \dodoi{10.1093/mnras/stz2410}
	
	\bibitem[{{Chabrier}(2003)}]{chabrier2003}
	{Chabrier}, G. 2003, \pasp, 115, 763, \dodoi{10.1086/376392}
	
	\bibitem[{{Costa} {et~al.}(2015){Costa}, {Sijacki}, \& {Haehnelt}}]{costa2015}
	{Costa}, T., {Sijacki}, D., \& {Haehnelt}, M.~G. 2015, \mnras, 448, L30,
	\dodoi{10.1093/mnrasl/slu193}
	
	\bibitem[{{Costa} {et~al.}(2014){Costa}, {Sijacki}, {Trenti}, \&
		{Haehnelt}}]{costa2014a}
	{Costa}, T., {Sijacki}, D., {Trenti}, M., \& {Haehnelt}, M.~G. 2014, \mnras,
	439, 2146, \dodoi{10.1093/mnras/stu101}
	
	\bibitem[{{da Cunha} {et~al.}(2013){da Cunha}, {Groves}, {Walter}, {Decarli},
		{Weiss}, {Bertoldi}, {Carilli}, {Daddi}, {Elbaz}, {Ivison}, {Maiolino},
		{Riechers}, {Rix}, {Sargent}, \& {Smail}}]{dacunha2013}
	{da Cunha}, E., {Groves}, B., {Walter}, F., {et~al.} 2013, \apj, 766, 13,
	\dodoi{10.1088/0004-637X/766/1/13}
	
	\bibitem[{{De Looze} {et~al.}(2014){De Looze}, {Cormier}, {Lebouteiller},
		{Madden}, {Baes}, {Bendo}, {Boquien}, {Boselli}, {Clements}, {Cortese},
		{Cooray}, {Galametz}, {Galliano}, {Graci{\'a}-Carpio}, {Isaak}, {Karczewski},
		{Parkin}, {Pellegrini}, {R{\'e}my-Ruyer}, {Spinoglio}, {Smith}, \&
		{Sturm}}]{delooze2014}
	{De Looze}, I., {Cormier}, D., {Lebouteiller}, V., {et~al.} 2014, \aap, 568,
	A62, \dodoi{10.1051/0004-6361/201322489}
	
	\bibitem[{{De Rosa} {et~al.}(2014){De Rosa}, {Venemans}, {Decarli}, {Gennaro},
		{Simcoe}, {Dietrich}, {Peterson}, {Walter}, {Frank}, {McMahon}, {Hewett},
		{Mortlock}, \& {Simpson}}]{derosa2014}
	{De Rosa}, G., {Venemans}, B.~P., {Decarli}, R., {et~al.} 2014, \apj, 790, 145,
	\dodoi{10.1088/0004-637X/790/2/145}
	
	\bibitem[{{De Vis} {et~al.}(2019){De Vis}, {Jones}, {Viaene}, {Casasola},
		{Clark}, {Baes}, {Bianchi}, {Cassara}, {Davies}, {De Looze}, {Galametz},
		{Galliano}, {Lianou}, {Madden}, {Manilla-Robles}, {Mosenkov}, {Nersesian},
		{Roychowdhury}, {Xilouris}, \& {Ysard}}]{devis2019}
	{De Vis}, P., {Jones}, A., {Viaene}, S., {et~al.} 2019, \aap, 623, A5,
	\dodoi{10.1051/0004-6361/201834444}
	
	\bibitem[{{Decarli} {et~al.}(2018){Decarli}, {Walter}, {Venemans},
		{Ba{\~n}ados}, {Bertoldi}, {Carilli}, {Fan}, {Farina}, {Mazzucchelli},
		{Riechers}, {Rix}, {Strauss}, {Wang}, \& {Yang}}]{decarli2018}
	{Decarli}, R., {Walter}, F., {Venemans}, B.~P., {et~al.} 2018, \apj, 854, 97,
	\dodoi{10.3847/1538-4357/aaa5aa}
	
	\bibitem[{{Di Mascia} {et~al.}(2023){Di Mascia}, {Carniani}, {Gallerani},
		{Vito}, {Pallottini}, {Ferrara}, \& {Valentini}}]{dimascia2022}
	{Di Mascia}, F., {Carniani}, S., {Gallerani}, S., {et~al.} 2023, \mnras, 518,
	3667, \dodoi{10.1093/mnras/stac3306}
	
	\bibitem[{{Di Mascia} {et~al.}(2021){Di Mascia}, {Gallerani}, {Behrens},
		{Pallottini}, {Carniani}, {Ferrara}, {Barai}, {Vito}, \&
		{Zana}}]{dimascia2021}
	{Di Mascia}, F., {Gallerani}, S., {Behrens}, C., {et~al.} 2021, \mnras, 503,
	2349, \dodoi{10.1093/mnras/stab528}
	
	\bibitem[{{Duras} {et~al.}(2017){Duras}, {Bongiorno}, {Piconcelli}, {Bianchi},
		{Pappalardo}, {Valiante}, {Bischetti}, {Feruglio}, {Martocchia}, {Schneider},
		{Vietri}, {Vignali}, {Zappacosta}, {La Franca}, \& {Fiore}}]{duras2017}
	{Duras}, F., {Bongiorno}, A., {Piconcelli}, E., {et~al.} 2017, \aap, 604, A67,
	\dodoi{10.1051/0004-6361/201731052}
	
	\bibitem[{{Eilers} {et~al.}(2022){Eilers}, {Simcoe}, {Yue}, {Mackenzie},
		{Matthee}, {Durovcikova}, {Kashino}, {Bordoloi}, \& {Lilly}}]{eilers2022}
	{Eilers}, A.-C., {Simcoe}, R.~A., {Yue}, M., {et~al.} 2022, arXiv e-prints,
	arXiv:2211.16261, \dodoi{10.48550/arXiv.2211.16261}
	
	\bibitem[{{Feruglio} {et~al.}(2018){Feruglio}, {Fiore}, {Carniani}, {Maiolino},
		{D'Odorico}, {Luminari}, {Barai}, {Bischetti}, {Bongiorno}, {Cristiani},
		{Ferrara}, {Gallerani}, {Marconi}, {Pallottini}, {Piconcelli}, \&
		{Zappacosta}}]{feruglio2018}
	{Feruglio}, C., {Fiore}, F., {Carniani}, S., {et~al.} 2018, \aap, 619, A39,
	\dodoi{10.1051/0004-6361/201833174}
	
	\bibitem[{{Foreman-Mackey} {et~al.}(2013){Foreman-Mackey}, {Hogg}, {Lang}, \&
		{Goodman}}]{foreman2013}
	{Foreman-Mackey}, D., {Hogg}, D.~W., {Lang}, D., \& {Goodman}, J. 2013, \pasp,
	125, 306, \dodoi{10.1086/670067}
	
	\bibitem[{{Galliano} {et~al.}(2018){Galliano}, {Galametz}, \&
		{Jones}}]{galliano2018}
	{Galliano}, F., {Galametz}, M., \& {Jones}, A.~P. 2018, \araa, 56, 673,
	\dodoi{10.1146/annurev-astro-081817-051900}
	
	\bibitem[{{Kashikawa} {et~al.}(2015){Kashikawa}, {Ishizaki}, {Willott},
		{Onoue}, {Im}, {Furusawa}, {Toshikawa}, {Ishikawa}, {Niino}, {Shimasaku},
		{Ouchi}, \& {Hibon}}]{kashikawa2015}
	{Kashikawa}, N., {Ishizaki}, Y., {Willott}, C.~J., {et~al.} 2015, \apj, 798,
	28, \dodoi{10.1088/0004-637X/798/1/28}
	
	\bibitem[{{Kimball} {et~al.}(2015){Kimball}, {Lacy}, {Lonsdale}, \&
		{Macquart}}]{kimball2015}
	{Kimball}, A.~E., {Lacy}, M., {Lonsdale}, C.~J., \& {Macquart}, J.~P. 2015,
	\mnras, 452, 88, \dodoi{10.1093/mnras/stv1160}
	
	\bibitem[{Kormendy \& Ho(2013)}]{kormendy2013}
	Kormendy, J., \& Ho, L.~C. 2013, Annual Review of Astronomy and Astrophysics,
	51, 511, \dodoi{10.1146/annurev-astro-082708-101811}
	
	\bibitem[{{Lagache} {et~al.}(2018){Lagache}, {Cousin}, \&
		{Chatzikos}}]{lagache2018}
	{Lagache}, G., {Cousin}, M., \& {Chatzikos}, M. 2018, \aap, 609, A130,
	\dodoi{10.1051/0004-6361/201732019}
	
	\bibitem[{{Li} {et~al.}(2022){Li}, {Venemans}, {Walter}, {Decarli}, {Wang}, \&
		{Cai}}]{li2022}
	{Li}, J., {Venemans}, B.~P., {Walter}, F., {et~al.} 2022, \apj, 930, 27,
	\dodoi{10.3847/1538-4357/ac61d7}
	
	\bibitem[{{Liu} {et~al.}(2022){Liu}, {Wang}, {Momjian}, {Wagg}, {Yang}, {An},
		{Shao}, {Carilli}, {Wu}, {Fan}, {Walter}, {Jiang}, {Li}, {Li}, {Fei}, \&
		{Xu}}]{liu2022}
	{Liu}, Y., {Wang}, R., {Momjian}, E., {et~al.} 2022, \apj, 929, 69,
	\dodoi{10.3847/1538-4357/ac5c50}
	
	\bibitem[{{Marconi} {et~al.}(2004){Marconi}, {Risaliti}, {Gilli}, {Hunt},
		{Maiolino}, \& {Salvati}}]{marconi2004}
	{Marconi}, A., {Risaliti}, G., {Gilli}, R., {et~al.} 2004, \mnras, 351, 169,
	\dodoi{10.1111/j.1365-2966.2004.07765.x}
	
	\bibitem[{{McMullin} {et~al.}(2007){McMullin}, {Waters}, {Schiebel}, {Young},
		\& {Golap}}]{mcmullin2007}
	{McMullin}, J.~P., {Waters}, B., {Schiebel}, D., {Young}, W., \& {Golap}, K.
	2007, in Astronomical Society of the Pacific Conference Series, Vol. 376,
	Astronomical Data Analysis Software and Systems XVI, ed. R.~A. {Shaw},
	F.~{Hill}, \& D.~J. {Bell}, 127
	
	\bibitem[{{Mortlock} {et~al.}(2011){Mortlock}, {Warren}, {Venemans}, {Patel},
		{Hewett}, {McMahon}, {Simpson}, {Theuns}, {Gonz{\'a}les-Solares}, {Adamson},
		{Dye}, {Hambly}, {Hirst}, {Irwin}, {Kuiper}, {Lawrence}, \&
		{R{\"o}ttgering}}]{mortlock2011}
	{Mortlock}, D.~J., {Warren}, S.~J., {Venemans}, B.~P., {et~al.} 2011, \nat,
	474, 616, \dodoi{10.1038/nature10159}
	
	\bibitem[{{Neeleman} {et~al.}(2021){Neeleman}, {Novak}, {Venemans}, {Walter},
		{Decarli}, {Kaasinen}, {Schindler}, {Ba{\~n}ados}, {Carilli}, {Drake}, {Fan},
		\& {Rix}}]{neeleman2021}
	{Neeleman}, M., {Novak}, M., {Venemans}, B.~P., {et~al.} 2021, \apj, 911, 141,
	\dodoi{10.3847/1538-4357/abe70f}
	
	\bibitem[{{Novak} {et~al.}(2019){Novak}, {Ba{\~n}ados}, {Decarli}, {Walter},
		{Venemans}, {Neeleman}, {Farina}, {Mazzucchelli}, {Carilli}, {Fan}, {Rix}, \&
		{Wang}}]{novak2019}
	{Novak}, M., {Ba{\~n}ados}, E., {Decarli}, R., {et~al.} 2019, \apj, 881, 63,
	\dodoi{10.3847/1538-4357/ab2beb}
	
	\bibitem[{{Pensabene} {et~al.}(2020){Pensabene}, {Carniani}, {Perna}, {Cresci},
		{Decarli}, {Maiolino}, \& {Marconi}}]{pensabene2020}
	{Pensabene}, A., {Carniani}, S., {Perna}, M., {et~al.} 2020, \aap, 637, A84,
	\dodoi{10.1051/0004-6361/201936634}
	
	\bibitem[{{Planck Collaboration} {et~al.}(2020){Planck Collaboration},
		{Aghanim}, {Akrami}, {Ashdown}, {Aumont}, {Baccigalupi}, {Ballardini},
		{Banday}, {Barreiro}, {Bartolo}, {Basak}, {Battye}, {Benabed}, {Bernard},
		{Bersanelli}, {Bielewicz}, {Bock}, {Bond}, {Borrill}, {Bouchet}, {Boulanger},
		{Bucher}, {Burigana}, {Butler}, {Calabrese}, {Cardoso}, {Carron},
		{Challinor}, {Chiang}, {Chluba}, {Colombo}, {Combet}, {Contreras}, {Crill},
		{Cuttaia}, {de Bernardis}, {de Zotti}, {Delabrouille}, {Delouis}, {Di
			Valentino}, {Diego}, {Dor{\'e}}, {Douspis}, {Ducout}, {Dupac}, {Dusini},
		{Efstathiou}, {Elsner}, {En{\ss}lin}, {Eriksen}, {Fantaye}, {Farhang},
		{Fergusson}, {Fernandez-Cobos}, {Finelli}, {Forastieri}, {Frailis},
		{Fraisse}, {Franceschi}, {Frolov}, {Galeotta}, {Galli}, {Ganga},
		{G{\'e}nova-Santos}, {Gerbino}, {Ghosh}, {Gonz{\'a}lez-Nuevo}, {G{\'o}rski},
		{Gratton}, {Gruppuso}, {Gudmundsson}, {Hamann}, {Handley}, {Hansen},
		{Herranz}, {Hildebrandt}, {Hivon}, {Huang}, {Jaffe}, {Jones}, {Karakci},
		{Keih{\"a}nen}, {Keskitalo}, {Kiiveri}, {Kim}, {Kisner}, {Knox},
		{Krachmalnicoff}, {Kunz}, {Kurki-Suonio}, {Lagache}, {Lamarre}, {Lasenby},
		{Lattanzi}, {Lawrence}, {Le Jeune}, {Lemos}, {Lesgourgues}, {Levrier},
		{Lewis}, {Liguori}, {Lilje}, {Lilley}, {Lindholm}, {L{\'o}pez-Caniego},
		{Lubin}, {Ma}, {Mac{\'\i}as-P{\'e}rez}, {Maggio}, {Maino}, {Mandolesi},
		{Mangilli}, {Marcos-Caballero}, {Maris}, {Martin}, {Martinelli},
		{Mart{\'\i}nez-Gonz{\'a}lez}, {Matarrese}, {Mauri}, {McEwen}, {Meinhold},
		{Melchiorri}, {Mennella}, {Migliaccio}, {Millea}, {Mitra},
		{Miville-Desch{\^e}nes}, {Molinari}, {Montier}, {Morgante}, {Moss}, {Natoli},
		{N{\o}rgaard-Nielsen}, {Pagano}, {Paoletti}, {Partridge}, {Patanchon},
		{Peiris}, {Perrotta}, {Pettorino}, {Piacentini}, {Polastri}, {Polenta},
		{Puget}, {Rachen}, {Reinecke}, {Remazeilles}, {Renzi}, {Rocha}, {Rosset},
		{Roudier}, {Rubi{\~n}o-Mart{\'\i}n}, {Ruiz-Granados}, {Salvati}, {Sandri},
		{Savelainen}, {Scott}, {Shellard}, {Sirignano}, {Sirri}, {Spencer},
		{Sunyaev}, {Suur-Uski}, {Tauber}, {Tavagnacco}, {Tenti}, {Toffolatti},
		{Tomasi}, {Trombetti}, {Valenziano}, {Valiviita}, {Van Tent}, {Vibert},
		{Vielva}, {Villa}, {Vittorio}, {Wandelt}, {Wehus}, {White}, {White},
		{Zacchei}, \& {Zonca}}]{planck2018}
	{Planck Collaboration}, {Aghanim}, N., {Akrami}, Y., {et~al.} 2020, \aap, 641,
	A6, \dodoi{10.1051/0004-6361/201833910}
	
	\bibitem[{{Schneider} {et~al.}(2015){Schneider}, {Bianchi}, {Valiante},
		{Risaliti}, \& {Salvadori}}]{schneider2015}
	{Schneider}, R., {Bianchi}, S., {Valiante}, R., {Risaliti}, G., \& {Salvadori},
	S. 2015, \aap, 579, A60, \dodoi{10.1051/0004-6361/201526105}
	
	\bibitem[{{Shao} {et~al.}(2019){Shao}, {Wang}, {Carilli}, {Wagg}, {Walter},
		{Li}, {Fan}, {Jiang}, {Riechers}, {Bertoldi}, {Strauss}, {Cox}, {Omont}, \&
		{Menten}}]{shao2019}
	{Shao}, Y., {Wang}, R., {Carilli}, C.~L., {et~al.} 2019, \apj, 876, 99,
	\dodoi{10.3847/1538-4357/ab133d}
	
	\bibitem[{{Tacconi} {et~al.}(2018){Tacconi}, {Genzel}, {Saintonge}, {Combes},
		{Garc{\'\i}a-Burillo}, {Neri}, {Bolatto}, {Contini}, {F{\"o}rster Schreiber},
		{Lilly}, {Lutz}, {Wuyts}, {Accurso}, {Boissier}, {Boone}, {Bouch{\'e}},
		{Bournaud}, {Burkert}, {Carollo}, {Cooper}, {Cox}, {Feruglio}, {Freundlich},
		{Herrera-Camus}, {Juneau}, {Lippa}, {Naab}, {Renzini}, {Salome}, {Sternberg},
		{Tadaki}, {{\"U}bler}, {Walter}, {Weiner}, \& {Weiss}}]{tacconi2018}
	{Tacconi}, L.~J., {Genzel}, R., {Saintonge}, A., {et~al.} 2018, \apj, 853, 179,
	\dodoi{10.3847/1538-4357/aaa4b4}
	
	\bibitem[{{Trakhtenbrot} {et~al.}(2017){Trakhtenbrot}, {Lira}, {Netzer},
		{Cicone}, {Maiolino}, \& {Shemmer}}]{trakhtenbrot2017}
	{Trakhtenbrot}, B., {Lira}, P., {Netzer}, H., {et~al.} 2017, \apj, 836, 8,
	\dodoi{10.3847/1538-4357/836/1/8}
	
	\bibitem[{{Tripodi} {et~al.}(2022){Tripodi}, {Feruglio, C.}, {Fiore, F.},
		{Bischetti, M.}, {D\'{}Odorico, V.}, {Carniani, S.}, {Cristiani, S.},
		{Gallerani, S.}, {Maiolino, R.}, {Marconi, A.}, {Pallottini, A.},
		{Piconcelli, E.}, {Vallini, L.}, \& {Zana, T.}}]{Tripodi2022}
	{Tripodi}, R., {Feruglio, C.}, {Fiore, F.}, {et~al.} 2022, A\&A, 665, A107,
	\dodoi{10.1051/0004-6361/202243920}
	
	\bibitem[{{Tsukui} {et~al.}(2023){Tsukui}, {Wisnioski}, {Krumholz}, \&
		{Battisti}}]{tsukui2023}
	{Tsukui}, T., {Wisnioski}, E., {Krumholz}, M.~R., \& {Battisti}, A. 2023, arXiv
	e-prints, arXiv:2302.07272, \dodoi{10.48550/arXiv.2302.07272}
	
	\bibitem[{{Valiante} {et~al.}(2011){Valiante}, {Schneider}, {Salvadori}, \&
		{Bianchi}}]{valiante2011}
	{Valiante}, R., {Schneider}, R., {Salvadori}, S., \& {Bianchi}, S. 2011,
	\mnras, 416, 1916, \dodoi{10.1111/j.1365-2966.2011.19168.x}
	
	\bibitem[{{Valiante} {et~al.}(2014){Valiante}, {Schneider}, {Salvadori}, \&
		{Gallerani}}]{valiante2014}
	{Valiante}, R., {Schneider}, R., {Salvadori}, S., \& {Gallerani}, S. 2014,
	\mnras, 444, 2442, \dodoi{10.1093/mnras/stu1613}
	
	\bibitem[{{Venemans} {et~al.}(2016){Venemans}, {Walter}, {Zschaechner},
		{Decarli}, {De Rosa}, {Findlay}, {McMahon}, \& {Sutherland}}]{venemans2016}
	{Venemans}, B.~P., {Walter}, F., {Zschaechner}, L., {et~al.} 2016, \apj, 816,
	37, \dodoi{10.3847/0004-637X/816/1/37}
	
	\bibitem[{{Venemans} {et~al.}(2017){Venemans}, {Walter}, {Decarli},
		{Ba{\~n}ados}, {Hodge}, {Hewett}, {McMahon}, {Mortlock}, \&
		{Simpson}}]{venemans2017a}
	{Venemans}, B.~P., {Walter}, F., {Decarli}, R., {et~al.} 2017, \apj, 837, 146,
	\dodoi{10.3847/1538-4357/aa62ac}
	
	\bibitem[{{Venemans} {et~al.}(2020){Venemans}, {Walter}, {Neeleman}, {Novak},
		{Otter}, {Decarli}, {Ba{\~n}ados}, {Drake}, {Farina}, {Kaasinen},
		{Mazzucchelli}, {Carilli}, {Fan}, {Rix}, \& {Wang}}]{venemans2020}
	{Venemans}, B.~P., {Walter}, F., {Neeleman}, M., {et~al.} 2020, \apj, 904, 130,
	\dodoi{10.3847/1538-4357/abc563}
	
	\bibitem[{{Volonteri}(2012)}]{volonteri2012}
	{Volonteri}, M. 2012, Science, 337, 544, \dodoi{10.1126/science.1220843}
	
	\bibitem[{{Walter} {et~al.}(2022){Walter}, {Neeleman}, {Decarli}, {Venemans},
		{Meyer}, {Weiss}, {Ba{\~n}ados}, {Bosman}, {Carilli}, {Fan}, {Riechers},
		{Rix}, \& {Thompson}}]{walter2022}
	{Walter}, F., {Neeleman}, M., {Decarli}, R., {et~al.} 2022, \apj, 927, 21,
	\dodoi{10.3847/1538-4357/ac49e8}
	
	\bibitem[{{Wang} {et~al.}(2019){Wang}, {Wang}, {Fan}, {Wu}, {Yang}, {Neri}, \&
		{Yue}}]{wang2019}
	{Wang}, F., {Wang}, R., {Fan}, X., {et~al.} 2019, \apj, 880, 2,
	\dodoi{10.3847/1538-4357/ab2717}
	
	\bibitem[{{Willott} {et~al.}(2015){Willott}, {Bergeron}, \&
		{Omont}}]{willott2015}
	{Willott}, C.~J., {Bergeron}, J., \& {Omont}, A. 2015, \apj, 801, 123,
	\dodoi{10.1088/0004-637X/801/2/123}
	
	\bibitem[{{Willott} {et~al.}(2017){Willott}, {Bergeron}, \&
		{Omont}}]{willott2017}
	---. 2017, \apj, 850, 108, \dodoi{10.3847/1538-4357/aa921b}
	
	\bibitem[{{Willott} {et~al.}(2013){Willott}, {Omont}, \&
		{Bergeron}}]{willott2013}
	{Willott}, C.~J., {Omont}, A., \& {Bergeron}, J. 2013, \apj, 770, 13,
	\dodoi{10.1088/0004-637X/770/1/13}
	
	\bibitem[{{Wu} {et~al.}(2015){Wu}, {Wang}, {Fan}, {Yi}, {Zuo}, {Bian}, {Jiang},
		{McGreer}, {Wang}, {Yang}, {Yang}, {Thompson}, \& {Beletsky}}]{wu2015}
	{Wu}, X.-B., {Wang}, F., {Fan}, X., {et~al.} 2015, \nat, 518, 512,
	\dodoi{10.1038/nature14241}
	
\end{thebibliography}

\begin{appendix}

\section{Posterior distributions}
	\label{app:a}

\noindent We model the dust continuum with a MBB function given by
 \begin{equation}\label{eq:MBB}
     S_{\nu_{\rm obs}}^{\rm obs} = S_{\nu/(1+z)}^{\rm obs} = \frac{\Omega}{(1+z)^3}[B_{\nu}(T_{\rm dust}(z))-B_{\nu}(T_{\rm CMB}(z))](1-e^{-\tau_{\nu}}),
 \end{equation}

\noindent where $\Omega = (1+z)^4A_{\rm gal}D_{\rm L}^{-2}$ is the solid angle with $A_{\rm gal}$, and $D_{\rm L}$ is the surface area and luminosity distance of the galaxy, respectively \citep{carniani+19}. The dust optical depth is
\begin{equation}
    \tau_{\nu}=\frac{M_{\rm dust}}{A_{\rm galaxy}}k_0\biggl(\frac{\nu}{250\ \rm GHz}\biggr)^{\beta},
\end{equation}
\noindent with $\beta$ the emissivity index and $k_0 = 0.45\  \rm cm^{2}\ g^{-1}$ the mass absorption coefficient \citep{beelen+2006}. The solid angle is estimated using the continuum mean size from resolved observations \citep{wang2019}. The effect of the CMB on the dust temperature is given by
\begin{equation}
    T_{\rm dust}(z)=((T_{\rm dust})^{4+\beta}+T_0^{4+\beta}[(1+z)^{4+\beta}-1])^{\frac{1}{4+\beta}},
\end{equation}
\noindent with $T_0 = 2.73$ K.
We also considered the contribution of the CMB emission given by $B_{\nu}(T_{\rm CMB}(z)=T_0(1+z))$ \citep{dacunha2013}. 

We model the low frequency radio emission using a PLCO that is 
\begin{equation}\label{eq:pl}
F_{\nu_{\rm rest}}=n\times (\nu_{\rm rest}/\nu_0)^{-\alpha}\times \exp(-\nu_{\rm rest}/\nu_{\rm cutoff})
\end{equation}

\noindent where $n$ is the normalization, $\alpha$ is the radio power-law spectral index, $\nu_0$ and $\nu_{\rm cutoff}$ are the reference frequency and the cutoff frequency, respectively. We set $\nu_0=59$ GHz to ease the fitting procedure and in order to minimize the covariance between $n$ and $\alpha$. This choice does not affect our results.

\begin{figure}
	\centering
	\includegraphics[width=0.9\linewidth]{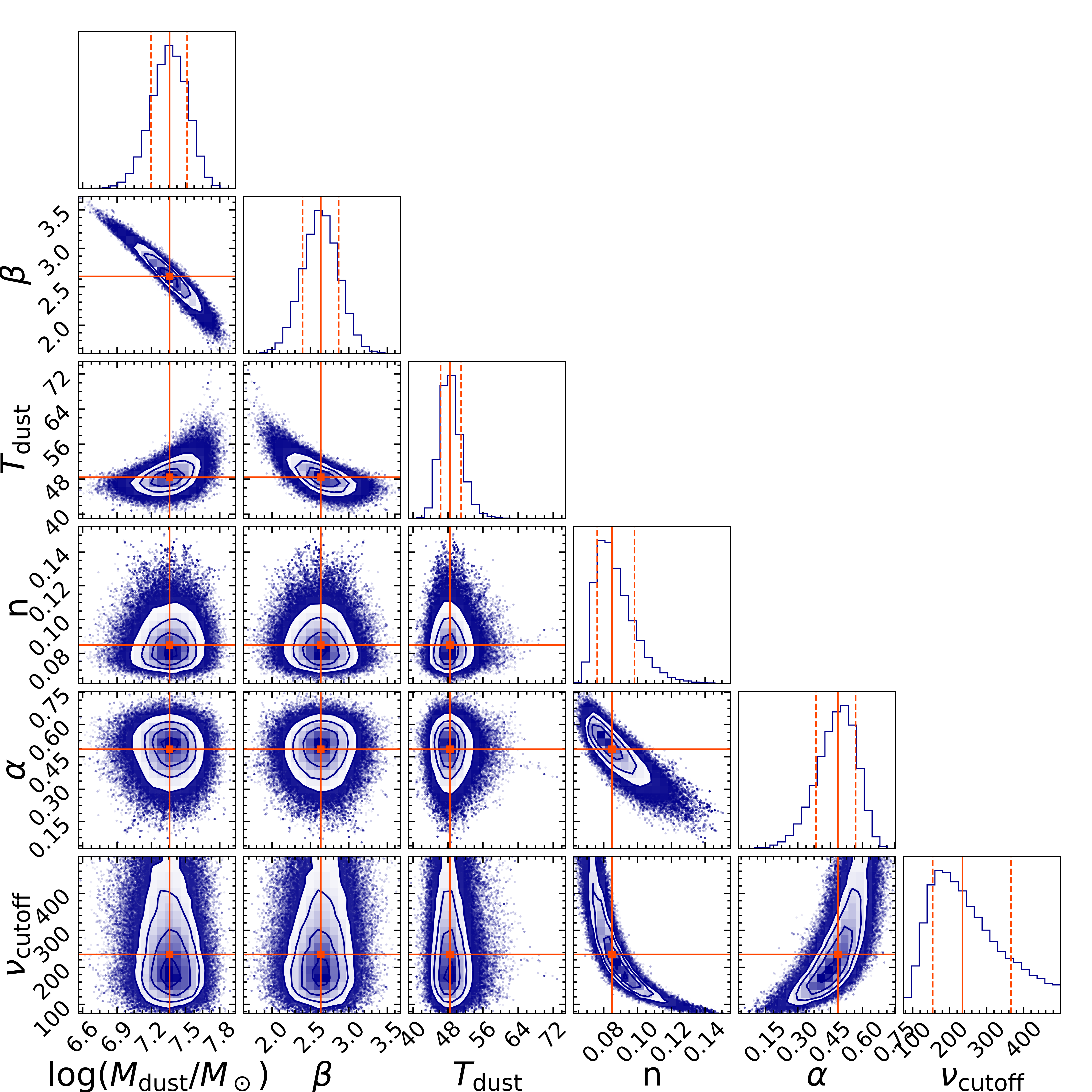}
\caption{\footnotesize Corner plot showing the six dimensional posterior probability distributions of $T_{\rm dust}, M_{\rm dust}, \beta,n,\alpha, \text{ and } \nu_{\rm cutoff}$. Orange solid lines on the posterior probability distributions indicate the best-fitting value for each parameter, while the dashed lines mark the 16\% and 84\% percentiles for each parameter.}
\label{fig:post-distributions}
\end{figure}

The total model has six fitting parameters: dust temperature ($T_{\rm dust}$), dust mass ($M_{\rm dust}$), $\beta$ entering in the MBB function, and $n$, $\alpha$, and $\nu_{\rm cutoff}$ for the PLCO. We explore the six dimensional parameter space using a Markov chain Monte Carlo (MCMC) algorithm implemented in the \texttt{EMCEE} package \citep{foreman2013}. We assume uniform priors for the fitting parameters: $10{\ \rm K} <T_{\rm dust}<300$ K, $10^{5} \ {\rm M_{\odot}}<M_{\rm dust}<10^{9}\ {\rm M_{\odot}}$, $1.0<\beta<3.0$, $0.001 {\ \rm mJy}<n<0.5$ mJy, $0.01 <\alpha<1.0$, $75.0{\ \rm GHz} <\nu_{\rm cutoff}<500$ GHz. The best-fit model has $T_{\rm dust}=48.4\pm2.3$ K, $M_{\rm dust}=(2.29\pm0.83)\times 10^7$ M$_\odot$, $\beta=2.63\pm 0.23$, $n=0.08\pm 0.01$ mJy, $\alpha=0.48\pm 0.09$ and $\nu_{\rm cutoff}=235 \pm 100$ GHz, obtained from a MCMC with 60 chains, 6000 trials and a burn-in phase of $\sim 100$. Figure \ref{fig:post-distributions} shows the posterior distributions of the six fitting parameters $T_{\rm dust}, M_{\rm dust}, \beta,n,\alpha, \text{ and } \nu_{\rm cutoff}$.

\end{appendix}

\end{document}